\documentclass[runningheads]{llncs}
\usepackage[T1]{fontenc}
\usepackage{graphicx}
\usepackage{booktabs}  
\usepackage{array}      
\usepackage{tabularx}

\usepackage{enumitem}   

\usepackage{ragged2e}   
\usepackage{multirow}   
\usepackage{float}
\usepackage[section]{placeins}

\begin{document}
\title{Memory-Based vs. Context-Only Conditioning Produces Distinct Behavioral Patterns in Stateful Personalization}
%
\titlerunning{Memory-Based vs. Context-Only Conditioning}
%
\author{Junsoo Park\orcidID{0009-0003-4450-0103} 
\and
Youssef Medhat \and
Htet Phyo Wai \and
Ploy Thajchayapong\orcidID{0009-0000-5993-1094} \and
Ashok K. Goel\orcidID{0000-0003-4043-0614}}
\authorrunning{J. Park et al.}
%
\institute{Georgia Institute of Technology, Atlanta, GA, USA\\
\email{jpark3232@gatech.edu, ymedhat3@gatech.edu, hwai6@gatech.edu, ploy@gatech.edu, ashok.goel@cc.gatech.edu}}
%

\maketitle              
\begin{abstract}
We study how conditioning context shapes personalization behavior in a teacher-facing educational recommender system. We compare contextual conditioning based on the current student question with memory-based conditioning using persistent learner information. Using deviation correlation and paired statistical tests, we find that contextual recommendations exhibit stronger question-level responsiveness, while memory-based recommendations exhibit history-dependent behaviors, including learner-specific differentiation under identical input. Teacher-facing evaluation signals suggest these recommendations are interpretable and actionable. These results indicate that embedding-based similarity metrics capture responsiveness to the current question but do not characterize personalization grounded in learner history, motivating behavior-level diagnostics for studying conditioning effects.
\keywords{Intelligent tutoring systems, Learner modeling, Stateful personalization, Memory-based conditioning, Large language models}
\end{abstract}

\section{Introduction}
Educational AI systems show promise for supporting student learning, yet instructors require interpretable engagement data and actionable instructional insights to personalize teaching effectively \cite{hershkovitz2024instructors,zou2025digital,bernacki2021systematic}. This has motivated teacher-facing AI systems that augment, rather than replace, instructional decision-making \cite{holstein2019complementarity}. We study a teacher-facing recommender system that generates instructional recommendations from student--AI question interactions, focusing on conditioning context as a central design variable. Recommendations may use contextual (episodic) conditioning based on the current student question or memory-based (stateful) conditioning using persistent learner information derived from prior interactions. This distinction reflects a broader shift from contextual responsiveness toward memory-based personalization. Here, memory-based conditioning refers to the use of interaction-derived learner information, without relying on identity-based attributes, as inputs to generation, consistent with emerging memory-augmented LLM approaches \cite{zhang2025survey,westhausser2025enabling}. We define a system as \emph{stateful} if recommendations depend on accumulated learner information rather than solely on the current interaction, enabling a controlled comparison between question-level responsiveness and history-dependent personalization not captured by existing embedding-based metrics. Throughout, we use \emph{contextual} and \emph{memory-based} as the primary conditioning terms, with \emph{question-conditioned} and \emph{learner-conditioned} used only for clarity. We use responsiveness to denote the extent to which recommendations vary with the current student question, as measured by embedding-based similarity and ask:

\textbf{RQ1:} How does memory-based conditioning change the personalization behaviors of a teacher-facing educational recommender compared to contextual conditioning?

\textbf{H1:} Memory-based conditioning produces qualitatively different personalization behaviors than contextual conditioning, enabling learner-specific differentiation under identical input.

\section{Background}
Personalized learning emphasizes adapting instructional support to learner needs over time \cite{bloom1968learning,tomlinson1999differentiated}. Adaptive educational systems operationalize this through persistent learner models that inform instructional decisions across interactions \cite{corbett1994knowledge,brusilovsky2001adaptive}. Recent teacher-facing AI systems extend this paradigm by surfacing interpretable signals from student interactions to support instructor decision-making \cite{hershkovitz2024instructors,zou2025digital}. Prior work has explored human--AI teaching systems leveraging bidirectional feedback \cite{park2025human} and system designs for AI-driven customization with teacher-in-the-loop refinement \cite{basu2025bidirectional}, motivating representation-based approaches grounded in interaction history. 

In parallel, large language model (LLM) systems have largely relied on contextual generation, where outputs depend on the current input or retrieved information at query time. Retrieval-augmented generation (RAG) improves grounding but remains episodic and query-dependent \cite{gao2023retrieval}. Emerging memory-augmented approaches instead introduce persistent representations of user history, enabling personalization across interactions rather than within a single episode \cite{zhang2025survey,westhausser2025enabling}. This reflects a shift from contextual responsiveness toward memory-based personalization, whose behavioral implications for educational recommender systems remain under-explored. Evaluating personalization in such systems is challenging because instructional recommendations often admit multiple valid forms and lack a single correct output. Embedding-based diagnostics, such as deviation correlation, capture episodic responsiveness but do not reflect personalization grounded in accumulated learner state. We therefore adopt Hattie and Timperley’s feedback framework \cite{hattie2007power} to structure instructional intent through \textit{Feed Up}, \textit{Feed Back}, and \textit{Feed Forward}. Finally, prior work cautions against identity-based personalization, which is difficult to operationalize and risks misrepresentation in automated systems \cite{bernacki2021systematic}. In contrast, this work focuses on memory-based personalization derived from interaction history and behavioral signals.

\section{Study Design}
\begin{figure}
\centering
\includegraphics[scale=0.5]{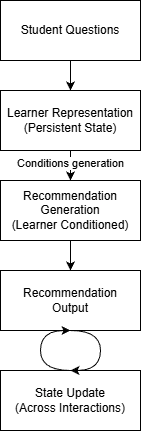}
\caption{Memory-based recommendation pipeline (learner-conditioned). The learner state functions as a persistent memory across interactions, retrieved at each step to guide pedagogical tactic selection prior to generation. The same generation mechanism is used across conditions; memory-based conditioning introduces memory retrieval and tactic selection.}
\label{fig:architecture}
\end{figure}

To address RQ1, we conduct a controlled comparison between two recommender configurations that share the same language model and generation procedure. The only manipulation is the conditioning information provided at generation time: the contextual (question-conditioned) system receives only the current student question, whereas the memory-based (learner-conditioned) system additionally receives a persistent learner state and a selected pedagogical tactic (Figure~\ref{fig:architecture}). In the memory-based configuration, the system retrieves a persistent learner representation consisting of need vectors and persona labels derived from prior interactions. These features guide pedagogical tactic selection prior to generation. The contextual baseline omits these inputs, ensuring that any observed differences arise solely from conditioning context.

\subsection{Data Collection}
We use anonymized interaction logs from an AI conversational agent deployed in graduate-level computer science courses at Georgia Institute of Technology \cite{goel2018jill}. The dataset contains 8{,}838 student questions from approximately 200 students across multiple course offerings. For controlled comparison, we randomly sample 50 questions and evaluate both conditions on identical inputs, differing only in conditioning context. Learner-state fields are derived from prior interactions and used solely as conditioning inputs. All data were anonymized prior to analysis and managed within a production data architecture for AI-augmented learning \cite{goel2025a4l,thajchayapong2025evolution}, supporting pseudo-anonymized learner representations while preserving interaction-level signals for memory-based conditioning. Instructor interviews were conducted under institutional review board (IRB) approval as part of ongoing system co-design.

\subsection{Conditioning and Experimental Setup}
The learner state summarizes prior interactions into interpretable fields used at generation time, including a need vector, dominant instructional need, and persona label. These features persist across interactions and are used solely as conditioning inputs to guide pedagogical tactic selection. In the memory-based (learner-conditioned) system, one of three pedagogical roles adapted from Hattie and Timperley’s framework \cite{hattie2007power}---\textit{Feed Up}, \textit{Feed Back}, or \textit{Feed Forward}---is selected using a fixed alignment heuristic that maps learner need profiles to instructional tactics. The selected tactic and learner-state fields are incorporated into the generation prompt.

We compare contextual and memory-based configurations that differ only in the conditioning inputs provided at generation time. Table~\ref{tab:prompt_skeletons} and Figure~\ref{fig:prompt_delta} illustrate the resulting prompt differences for the same question question input and define how learner information is incorporated into generation.

\begin{table}
\centering
\small
\caption{Abridged prompt skeletons illustrating the conditioning delta (same student question).}
\label{tab:prompt_skeletons}
\begin{tabular}{p{2.9cm}p{8.0cm}}
\toprule
\textbf{Condition} & \textbf{Abridged prompt inputs} \\
\midrule
Contextual (question-conditioned) &
\texttt{Question: \{student\_question\} $\rightarrow$ Generate teacher-facing recommendation.} \\
\midrule
Memory-based (learner-conditioned) &
\texttt{Learner state: \{top\_need, persona\}; Tactic: \{feed\_up/feed\_back/feed\_forward\}; Question: \{student\_question\} $\rightarrow$ Generate recommendation consistent with tactic.} \\
\bottomrule
\end{tabular}
\end{table}

\begin{figure}
\centering
\small
\setlength{\fboxsep}{6pt}
\fbox{
\begin{minipage}{0.95\columnwidth}
\raggedright
\textbf{Filled prompt examples (same student question)}

\vspace{0.5em}
\textbf{Contextual prompt (question-conditioned baseline)}

\footnotesize
Given a student question:

\emph{``True or false, design according to some people, is the cognitive task with the most economic value''}

Generate a teacher-facing pedagogical recommendation personalized for the student.\\

\vspace{0.75em}
\textbf{Memory-based prompt (learner-conditioned; added fields shown)}
\\
Student context: persona=\emph{Help Seekers}; top\_need=\emph{engagement};\\
Need vector: performance=0.38, engagement=0.62, skill\_progression=0.56;\\
Tactic: \emph{Feed Forward: Action Steps}.\\
Student question: \emph{``True or false, design...''}

[Directives and constraints omitted for brevity.]\\
\end{minipage}}
\caption{Prompt-level conditioning contrast for the same question. Memory-based conditioning introduces persistent learner state and a retrieved pedagogical tactic, functioning as memory inputs that guide generation.}
\label{fig:prompt_delta}
\end{figure}

\subsection{Analysis and Metrics}
We analyze personalization behavior through learner-specific divergence under identical input. To characterize question-level responsiveness, we use deviation correlation, defined as the correlation between deviations of question and recommendation embeddings from the classroom mean. Higher values indicate stronger responsiveness to the current question under episodic conditioning. Because this metric depends only on question and recommendation embeddings, it captures responsiveness to the current input rather than behavior driven by learner history.

\section{Results}

\subsection{Deviation-Based Similarity and Statistical Comparison}
We examine whether deviation-based diagnostics distinguish between conditioning contexts. This metric favors systems whose outputs are more responsive to the current question. Contextual recommendations exhibit higher deviation correlation ($\rho = 0.689$) than memory-based recommendations ($\rho = 0.578$), indicating stronger question-level responsiveness under contextual conditioning. Paired comparisons across matched interactions using t-tests and Wilcoxon signed-rank tests show that memory-based conditioning yields a statistically significant reduction in deviation correlation ($p \ll 0.01$) with large effect sizes, indicating systematic divergence from question-level responsiveness toward history-dependent behavior.

\subsection{Personalized Recommendations Under Identical Input}
To isolate learner-specific effects, we compare recommendations generated for different learners given the \emph{same} student question. Under memory-based conditioning, recommendations diverge because generation depends on stored learner information rather than on question content alone. Table~\ref{tab:divergence} shows a representative example in which two learners with different dominant instructional needs receive different recommendations despite identical input. Dominant instructional needs are derived from learner-state estimators and are treated here as inputs to the conditioning manipulation rather than as evaluation targets.

\begin{table}
\centering
\small
\caption{Personalized recommendations under identical input. Memory-based outputs diverge across learners due to differences in persistent learner state, illustrating memory-based personalization beyond contextual conditioning.}
\label{tab:divergence}
\begin{tabular}{p{3.3cm}p{4cm}p{4cm}}
\toprule
& \textbf{Learner A} & \textbf{Learner B} \\
\midrule
Dominant need & Engagement & Performance \\
Selected tactic & Feed Up & Feed Back \\
Instructional focus & Goal clarity and relevance & Mastery status and accuracy feedback \\
\midrule
Contextual output & \multicolumn{2}{p{7.7cm}}{\raggedright
Have the student justify their True/False choice using a claim--evidence--reasoning structure and rewrite the statement to make it testable.} \\
\midrule
Memory-based output & \raggedright
Clarify the purpose of the task and why evaluating claims about cognitive tasks matters in the course context; ask the student to restate the goal in their own words before proceeding. & \raggedright
Provide accuracy-based feedback relative to course mastery thresholds, identify the specific misconception in the response, and highlight concepts already mastered. \\
\end{tabular}
\end{table}

\subsection{Teacher-Facing Evaluation Signals}
To complement behavioral analysis, we incorporate lightweight teacher-facing evaluation signals from ongoing system co-design. Teaching fellows ($n=5$) rated perceived usefulness and usability using 5-point Likert-scale items adapted from established surveys \cite{brooke1996sus,davis1989technology}, with mean ratings exceeding 4.0 for both dimensions. Semi-structured interviews further highlighted the value of recommendation views for identifying intervention opportunities and interpreting student engagement patterns. These signals do not constitute outcome evaluation but indicate that the recommendations are interpretable and actionable, supporting the system’s practical relevance.

\section{Discussion}

\subsection{Episodic Versus Stateful Personalization}
Memory-based conditioning produces behavior that diverges from question-level responsiveness, reflecting the influence of accumulated learner information rather than the current query alone. We do not interpret this difference as inherently better or worse for learning outcomes. Instead, it reflects a shift in the type of system behavior being expressed: contextual conditioning emphasizes question-level responsiveness, whereas memory-based conditioning incorporates learner history. This distinction is analogous to differences between shallow and deep similarity in knowledge-based AI \cite{goel2025kbai}. The observed divergence for the same question input indicates that memory-based conditioning induces behavior that cannot be explained by context alone, suggesting a transition from episodic responsiveness to stateful personalization.

\subsection{Implications For Evaluating Personalization}
Embedding-based similarity metrics such as deviation correlation capture question-level responsiveness but do not reflect personalization grounded in learner history. Evaluating personalization therefore requires distinguishing between types of behavior reflected by similarity metrics and incorporating complementary analyses of learner-specific differentiation. These effects also depend on the informativeness of interaction data used to construct learner representations. Complementary work evaluates the distinctiveness of such representations as a structural diagnostic of representation quality \cite{park2026evaluating}, providing a representation-level perspective alongside the behavior-level analysis in this work.

\subsection{Implications For System Design and Deployment}
Conditioning context functions as a design variable that determines the range of personalization behaviors a recommender can exhibit. Providing persistent learner information at generation time enables differentiation for the same question input without modifying the language model or training procedure. While this study uses a limited sample of interactions, larger datasets would improve stability and enable analysis of longer-term patterns. Finally, in contrast to identity-based personalization—which is difficult to operationalize and risks misrepresentation \cite{bernacki2021systematic}—this work focuses on memory-based personalization derived from interaction history. The underlying data architecture supports this distinction by maintaining pseudo-anonymized learner representations \cite{goel2025a4l,thajchayapong2025evolution}, providing a practical boundary between memory-based and identity-driven approaches.

\section{Conclusion}
RQ1 examined how conditioning on persistent learner state changes personalization behavior in a teacher-facing educational recommender system. Holding the language model and generation procedure fixed, we show that adding learner history at generation time shifts personalization from contextual, question-level responsiveness to memory-based, state-dependent instructional differentiation. In particular, memory-based conditioning enables different recommendations for different learners under identical question input through tactic selection informed by learner state. These findings support H1: conditioning on persistent learner state produces qualitatively different personalization behaviors than contextual conditioning alone. We further show that deviation correlation favors contextual systems because it captures how closely recommendations align with the current question, but does not reflect behavior grounded in learner history. We do not claim this difference implies improved learning outcomes. Rather, it reflects a shift in the underlying behavior: contextual conditioning produces responses tightly coupled to the current question, whereas memory-based conditioning incorporates structure derived from prior interactions. This shift can be interpreted as moving from episodic responses driven by surface-level similarity metrics to structured behavior informed by prior interactions, analogous to shallow versus deep similarity in knowledge-based AI \cite{goel2025kbai}. These results highlight that evaluating personalization requires analyzing behavioral patterns rather than relying on a single similarity metric. This work proposes behavior-level diagnostics for studying how conditioning context shapes system outputs. Complementary work on representation distinctiveness \cite{park2026evaluating} evaluates the structure and separability of learner-state representations. These approaches enable analysis of both representation quality and behavioral effects, providing a practical approach for analyzing personalization behavior without outcome labels.

\subsubsection{Acknowledgements} We thank members of the A4L team in Georgia Tech's Design Intelligence Lab (DILab; dilab.gatech.edu) for their contributions to this research. This research is supported by a US National Science Foundation grant (Grant No. 2247790) to the National AI Institute for Adult Learning and Online Education (AI-ALOE; aialoe.org).


%
%
%
\bibliographystyle{splncs04}
\bibliography{refs}
%





\end{document}